# Incipient ionic conductors: Ion-constrained lattices achieving superionic-like thermal conductivity by extreme anharmonicity

## Authors

Yongheng Li[1], Chunqiu Lu[2], Bin Wei[3#], Cong Lu[4], Xingang Jiang[1], Daisuke Ishikawa[5], Taishun Manjo[5], Caofeng Pan[6#], Alfred Q. R. Baron[5], Jiawang Hong[1#]

## Affiliations

1. School of Aerospace Engineering, Beijing Institute of Technology, Beijing, 100081, China

2. School of Materials Science and Engineering, Nanyang Technological University, 639798, Singapore

3. School of Materials Science and Engineering, Henan Polytechnic University, Jiaozuo 454000, China

4. Multi-disciplinary Research Division, Institute of High Energy Physics, Chinese Academy of Sciences, Beijing, 100049, China

5. Materials Dynamics Laboratory, RIKEN SPring-8 Center, RIKEN, 1-1-1 Kouto, Sayo, Hyogo 679-5148, Japan

6. Institute of Atomic Manufacturing, Beihang University, Beijing 100191, China

Corresponding authors: binwei@hpu.edu.cn, pancaofeng@buaa.edu.cn, jw_hong@bit.edu.cn

## Abstract

Phonon liquid-like thermal conduction in the solid state enables superionic conductors to serve as efficient thermoelectric device candidates. While liquid-like motion of ions effectively suppresses thermal conductivity ($\kappa$), their high mobility concurrently triggers material degradation due to undesirable ion migration and consequent metal deposition, making it still a challenge to balancing low $\kappa$ and high stability. Here, we report a superionic-like thermal transport alongside restricted long-range ion migration in $CsCu_2I_3$ with incipient ionic conduction, using synchrotron X-ray diffraction, inelastic X-ray scattering, and machine-learning potential-based simulations. We reveal that the Cu ions exhibit confined migration between $CuI_4$ tetrahedra at high temperatures, displaying extreme anharmonicity of dominated phonons beyond conventional rattling and comparable to that in superionic conductorsl. Consequently, a glass-like $\kappa$ (~0.3 W $m^{-1}$ $K^{-1}$ at 300 K) following the relationship of $\kappa \sim T^{0.17}$, was achieved along the *x*-direction, where Cu ion migration is three oders of magnitude lower than in superionic conductors. These results highlight the advantage of incipient ionic conductors in simultaneously maintaining both low $\kappa$ and high stability, elucidate the thermal transport mechanism via ion migration constraints, and pave an effective pathway toward ultralow thermal conductivity in ionic conductors.

## Introduction

Low thermal conductivity, $\kappa$, is of special interest and highly desirable in thermal insulation, thermodiodes, and thermoelectrics, etc., but it cannot be arbitrarily low due to the intrinsic limitation by atomic vibrations in most semiconductors and insulators[1,2]. To overcome this issue intrinsically in materials such as thermoelectric, substantial progress has been made to suppress phonon propagation via structural design strategies from "phonon-glass" to "phonon-liquid"[3]. Superionic conductors, by exploiting the liquid-like transport of phonons, have successfully realized such concepts, achieving ultralow $\kappa$ to a glass limit in the solid state. For example, in $Cu_2Se$, the random distribution of Cu ions allows them to migrate freely, resulting in strong scattering of phonons and thus low $\kappa$ (< 0.5 W $m^{-1}$ $K^{-1}$) over a wide temperature range[4]. In $CuCrSe_2$, specific phonons dominated by Cu ions break down due to giant anharmonicity and disorder, yielding low $\kappa$ (< 1 W $m^{-1}$ $K^{-1}$)[5]. In $Ag_8SnSe_6$, mobile Ag ions with high mobility ($\sigma$ ~ 0.1-10 S $m^{-1}$) dominated overdamped phonons reveals extreme anharmonicity, responsible for low $\kappa$ (< 0.5 W $m^{-1}$ $K^{-1}$) and fast diffusion[6]. However, superionic conductors suffer from material degradation under operating conditions due to major ion migration and metal decomposition caused by electric fields or thermal gradients, hindering their practical application[7]. Thus, outstanding questions arising from such issue remain on how to suppress ion migration to realize high stability while maintaining liquid-like thermal conductivity in ionic conductors.

Recent studies have proposed the concept of incipient ionic conductors, materials poised at the threshold of ionic conduction, where ions exhibit substantial local mobility yet experience suppressed long-range diffusion due to structural confinement within the host lattice[8]. This unique state bridges conventional electrical and typical supurionic conductors, where collective ion motions have certain impact on heat transfer without long-range displacement. As observed in tetrahedrite $Cu_{12}Sb_4S_{13}$, Cu ions are confined to migrate (low mobility) only between the corners of a given $SCu(2)_6$ octahedra, which is beyond rattling and yields a low $\kappa$ (<1 W $m^{-1}$ $K^{-1}$) even in the pristine system with crystalline stability[8,9]. As exemplarily discussed above, the heat transport mechanisms are significantly different in superionic conductors, such as ionic diffusion induced phonon (selectively) broken down in $Ag(Cu)CrSe_2$[5,10], extremely phonon anharmonicity in $Ag_8SnSe_6$[6], and no ionic/phonon conductivity correlation in Ag Argyrodites[11]. Inspired by this, seeking new systems with constrained ionic migration is crucial to further validate the incipiency, ultimately revealing the fundamental link between lattice dynamics and ionic transport in confined frameworks.

Possessing soft lattice structure and element synergy effect, ternary metal halides have been recently reported to suggest their potential for thermoelectric applications due to the ultralow $\kappa$ and tunable electronic properties.[12] In alkali-metal halides containing Cu and Ag, $(AX)_x$–$(MX)_{1-x}$ (A = Cu/Ag, M = K/Rb/Cs, and X = Cl/Br/I), the halide polyhedral framework enables certain freedom of Cu/Ag to large motion, resulting in relatively low ionic conductivity ($\sigma$ < $10^{-6}$ S $m^{-1}$ at room temperature), making them appropriate candidates for studing incipient ionic behavior in thermal transport[13]. Here, we selected $CsCu_2I_3$ as a demo to investigate this issue through a combination of synchrotron X-ray diffraction, inelastic X-ray scattering, and machine-learning potential-based molecular dynamics (MD) simulations. We reveal that the migration of Cu ions is predominantly confined to the approximate *x*-direction via an exchange mechanism between neighboring $CuI_4$ tetrahedra. This restricted ion motion induces extreme lattice anharmonicity and leading to the glaas-like $\kappa$ (~0.3 W $m^{-1}$ $K^{-1}$ at 300 K). Notably, the increase in thermal conductivity at high temperatures remains minimal, suggesting the mobility of ions has limited impact on thermal transport. This work not only establishes halide compounds as a promising platform for incipient ionic transport but also offers a novel design strategy for achieving ultralow thermal conductivity.

## Results and discussion

### Constrained migration of Cu ions

High-resolution single crystal synchrotron X-ray diffraction (SCSXRD) data were collected at beamline BL02B1 (SPring-8) to probe the temperature-dependent structural evolution of $CsCu_2I_3$, detailed in Supplementary Fig. S1 and Tables S1-S3. As illustrated in Fig. 1a, the 300 K crystal structure adopts orthorhombic symmetry (space group *Cmcm*), where Cu ions occupy interstitial sites within an I anions tetrahedral sublattice, forming discrete $CuI_4$ units. These paired $CuI_4$ strings are confined within rhombic columnar cages of Cs cations and extend infinitely through edge-sharing, constituting the characteristic $Cu_2I_3$ chains. Structure factors extracted from SCSXRD enable the calculation of maximum-entropy method (MEM)[14,15], which reconstructs probable electron density (ED) distributions, providing a robust approach for analyzing atomic localization[16], disorder[17,18], and ion migration intermediate (MID) pathways[19]. Fig. 1b reveals localized ED peaks at adjacent Cu sites below 500 K, whereas a continuous diffusion channel emerges at 500 K, evidenced by connected ED maxima along MID in normalized profiles (Fig. 1c). This channel suggests possible Cu ion diffusion or hopping between neighboring sites above 300 K. However, the low ED intensity within the channel (maximum ED < ~4% of Cu-site peaks) indicates restricted ion mobility. Consequently, $CsCu_2I_3$ can be classified as an incipient ionic conductor with thermally activated yet kinetically limited ion transport.

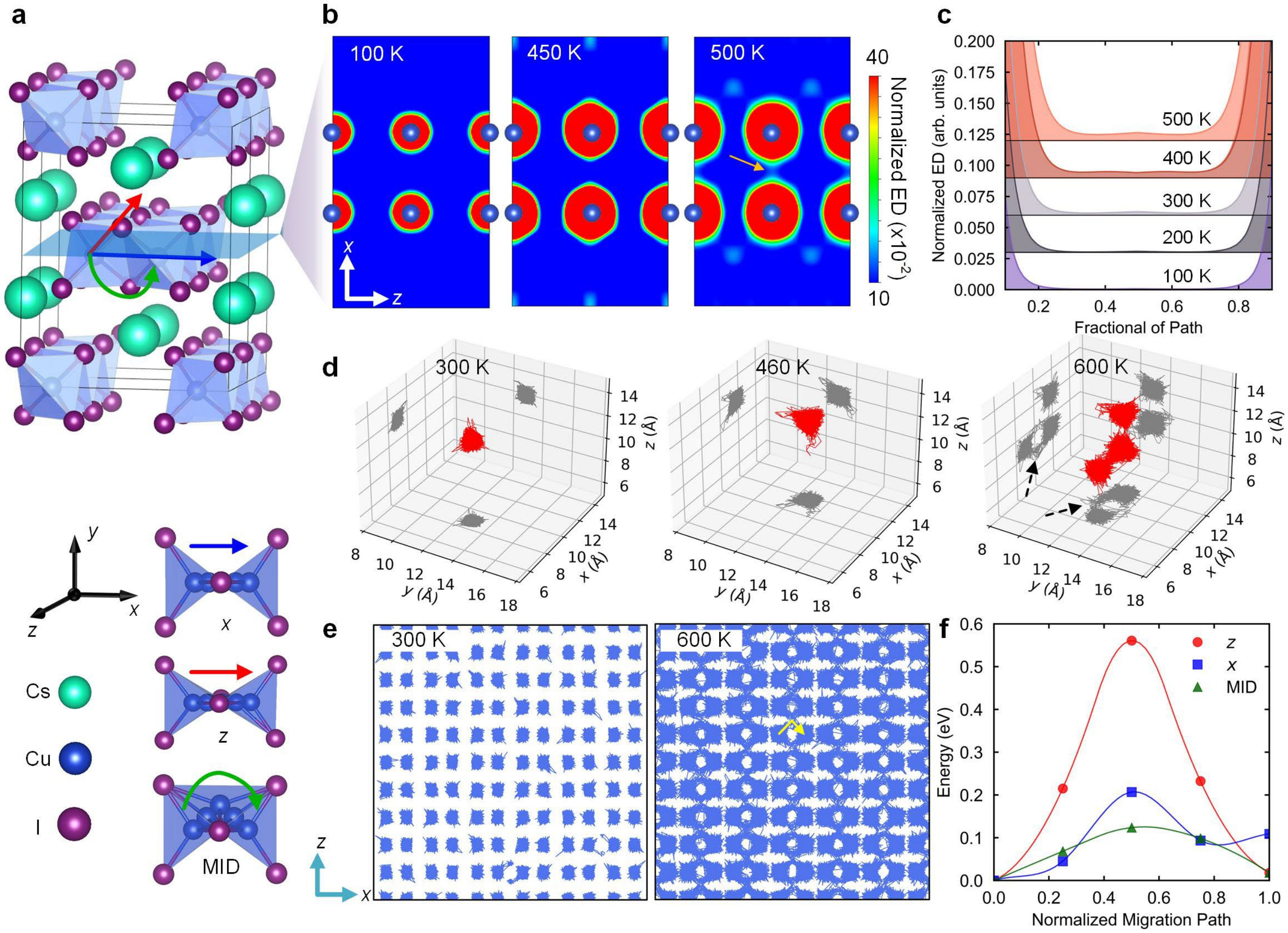

**Fig. 1 | Temperature-driven motion of Cu ions in $CsCu_2I_3$. a** Crytal structure of $CsCu_2I_3$ with the (010) plane shadowed in light blue and migration pathways of Cu ions along *x*- ([100]), *z*- ([001]), and intermediate (MID) directions indicated by blue, red, and green arrows, respectively. **b** Normalized electron density (ED) maps on the (100) plane  from structure factors measured by single crystal synchrotron X-ray diffraction at 300, 450, and 500 K.

The orange arrow indicates the delocalization of Cu ions at 500 K. **c** Normalized ED profile through the MID pathways marked with green arrows in (**a**) with increasing temperature. **d** Trajectories of selected single Cu ions from molecular dynamics (MD) simulation at 300, 450, and 600 K, showing the exchange of Cu ions across neighboring sites, as intuitively exhibited by extended trajectories in the *xz*-plane at 300 K and 600 K (yellow arrow in **e**). **f** Calculated energy barrier along *x*-, *z*-, and MID-directions.

Machine-learning potential-based MD simulations reveal Cu ion hopping dynamics at elevated temperatures (Fig. 1d and Supplementary Fig. S2). At 300 K, trajectories exhibit spherical confinement, whereas anisotropic displacements emerge at 450 K, manifesting as tentacle-like extensions from the spherical profile. The observed geometry shows broader $Cu^+$ displacement amplitudes at 450 K versus 300 K. Upon heating to 600 K, explicit ion hopping events occur with three-dimensional characteristics: trajectories extend not only within the *xy*-plane but also sometimes along the *z*-direction, as visualized in the *xz*-plane projection (Fig. 1e). This behavior aligns with the continuous ED channels (Fig. 1b-c). The hopping mechanism primarily involves exchange processes between adjacent $CuI_4$ tetrahedra as yellow arrows shown in Fig. 1e. Energy barrier simulations further elucidate this phenomenon (Fig. 1f). The migration barrier along the MID pathway (0.12 eV) is significantly lower than along the *z*-direction (0.56 eV) or *x*-direction (0.21 eV). This energetic preference explains: (1) the formation of continuous ED channels specifically along MID pathways, (2) the prevalence of corner-sharing connections between Cu sites, and (3) the absence of direct inter-tetrahedral hopping along Cartesian axes.

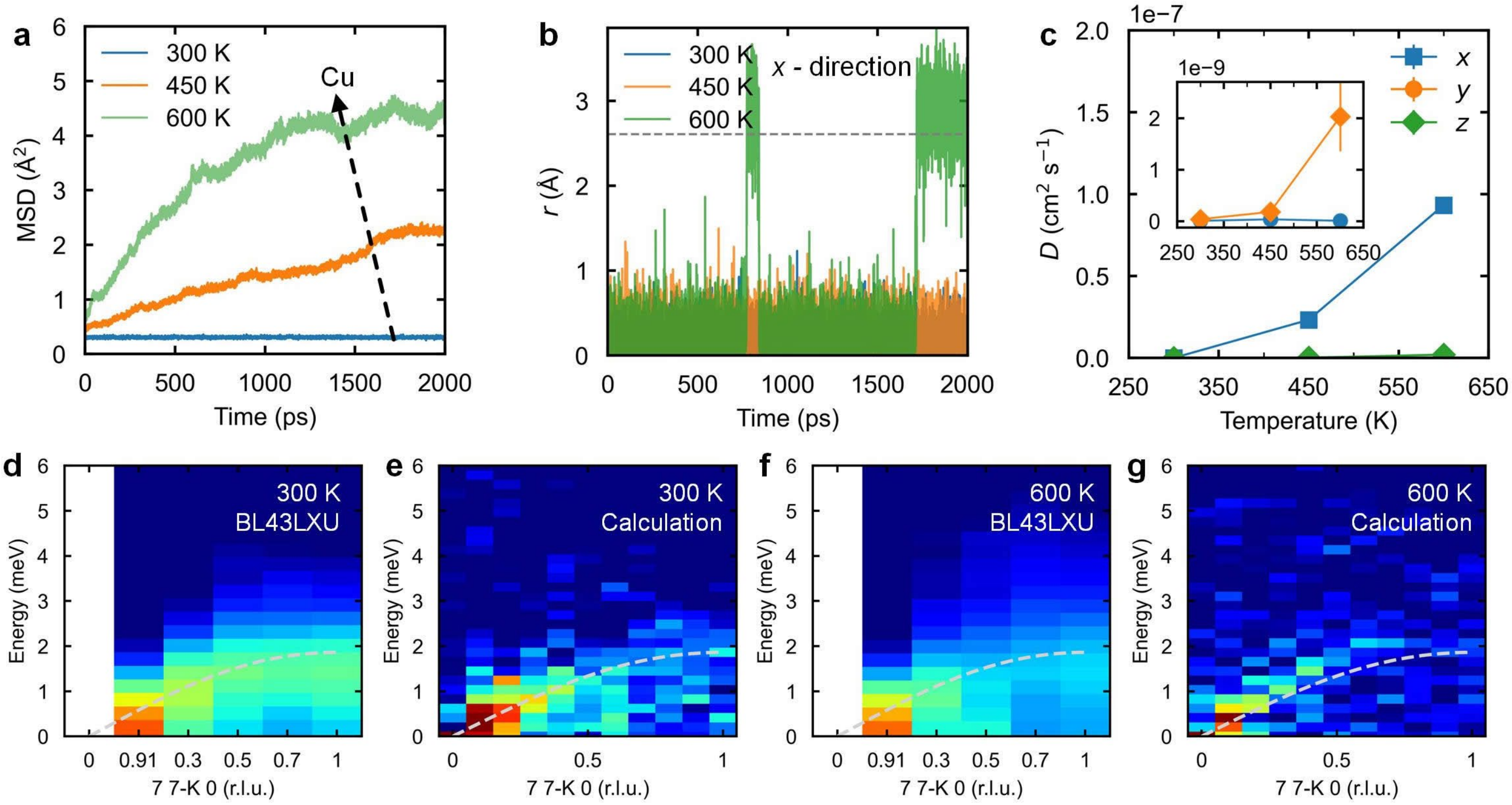

**Fig. 2 | Hopping of Cu ions and well-defined phonons in $CsCu_2I_3$. a** Mean square displacements (MSD) of Cu ions at different temperatures from molecular dynamics (MD) simulations. **b** Time-dependent displacement of a selected Cu ion along *x*-direction from MD. Grey dashed line indicates the distance between the nearest-neighboring $CuI_4$ tetrahedra. **c** Diffusion coefficient of Cu ions along different directions at different temperatures. Inset shows an enlarged view of the diffusivity between 0 and $2.5\times10^{-9}$ $cm^2$ $s^{-1}$, illustrating diffusivity along the *y*- and *z*- directions. (**d**) and (**f**) are measured phonon dispersion along the [0 K 0] direction at Q = (7 7 0), while (**e**) and (**g**) are simulated phonon dispersion along the same direction at Q = (7 7 0) at 300 K and 600 K, respectively. The dashed lines in (d-f) are calculated TA branch at 300 K.

Cu ion hopping dynamics can be characterized by time-dependent mean square displacement (MSD) analysis (Fig. 2a). At 300K, MSD of $Cu^+$ remains temporally invariant, indicating localized oscillations. Elevated temperatures induce nonlinear MSD evolution, with pronounced curvature emerging at 600 K, signaling frequent inter-tetrahedral hopping between adjacent $CuI_4$ units. A projected time-resolved displacement trajectories further confirm the constrained hopping of selected Cu ions: at 300 K and 450 K, ions oscillate near equilibrium positions along Cartesian axes. At 600 K, hopping to adjacent $CuI_4$ tetrahedra occurs with directional preference—the hopping frequency along the *x*-direction exceeds that along the *z*-direction, and along the *y*-direction exhibits only oscillation (no hopping) (Fig. 2b vs. Supplementary Fig. S3). The ionic diffusion coefficient with high anisotropy agrees well with this directional preference of Cu ions (Fig. 2c). Diffusivities along all axes approach zero at 300 K and 460 K. At 600 K, the *x*-direction diffusivity peaks at $9.3 \times 10^{-8}$ $cm^2$ $s^{-1}$, while *z*-direction diffusivity reaches $2.0 \times 10^{-9}$ $cm^2$ $s^{-1}$—an order of magnitude lower. The *y*-direction diffusivity remains negligible ($< 10^{-9}$ $cm^2$ $s^{-1}$). Notably, hopping is restricted to nearest-neighbor tetrahedral exchanges, with only small probability of sequential hops to next-nearest sites, indicating suppressed long-range diffusion. This constraint is remarkable by comparing with other superionic conductors such as $1\times10^{-5}$ $cm^2$ $s^{-1}$ at 600 K in $AgCrSe_2$[20], $1.3\times10^{-5}$ $cm^2$ $s^{-1}$ at 500 K in $Cu_7PS_6$[21]. These results demonstrate an excellent agreement with the suppressed long-range ionic diffusion and previously reported low ionic conductivity[13], suggesting $CsCu_2I_3$ to serve as an appropriate incipient ionic conductor.

**Hopping Cu ions induced extreme phonon anharmonicity**

Figure 2d-g shows the measured phonon dispersions along $\Gamma$−Y (*y*) direction at 300 K and 600 K by inelastic X-ray scattering (IXS) at beamline BL43LXU (SPring-8), which are well reproduced by our simulation. The discrepancies in distribution of intensity is due to the smoothing functions applied in the calculations. Importantly, the transverse acoustic (TA) branch, typically employed to assess ionic diffusion, persist at elevated temperatures despite Cu ion hopping that drives the system beyond rattler-dominated dynamics. Although increased diffusivity is observed near the zone boundary at 600 K (attributable to thermally enhanced anharmonicity), the TA modes remain well-defined. Identical behavior is confirmed along $\Gamma$−X and $\Gamma$−Z directions (Supplementary Fig. S4), demonstrating thermal robustness. This phenomenon arises from constrained Cu ion migration, where the residence time ($\tau$) at each site substantially exceeds the phonon oscillation period[22]. Consequently, TA phonons propagate with crystalline coherence, which is a characteristic signature of incipient ionic conductors as well.

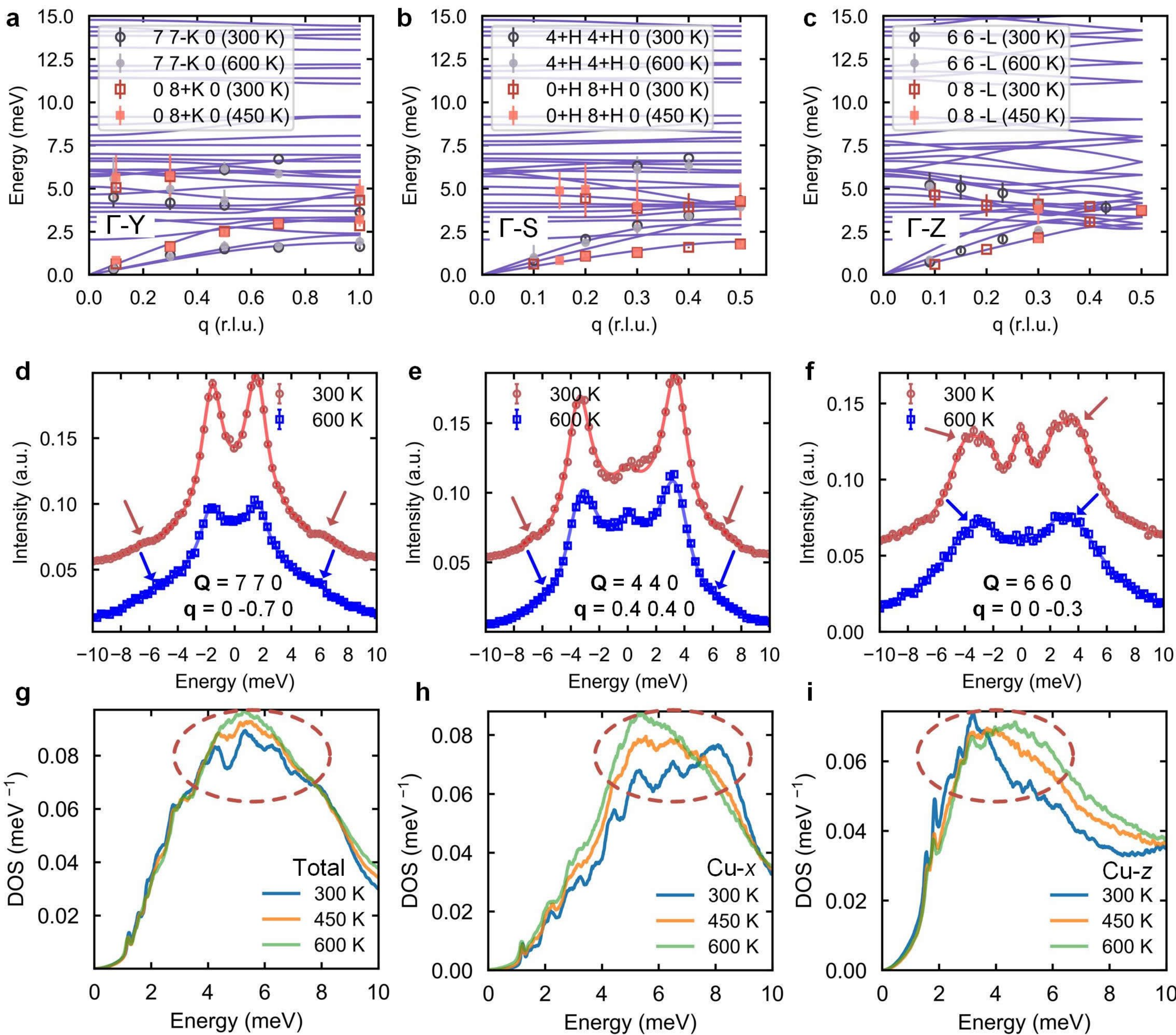

**Fig. 3 | Phonon dispersion and density of states in $CsCu_2I_3$.** (**a−c**) are measured phonon dispersion by inelastic X-ray scattering (IXS) along the $\Gamma$−Y, $\Gamma$−S and $\Gamma$−Z, respectively, overlaid with the calculated phonon dispersion at 300 K. Hollow markers, solid squares, and solid circles represent results at 300, 450, and 600 K. Error bar indicates the phonon linewidth. (**d−f**) are experimental constant-Q scans at Q = (7, 6.3, 0), (4.4, 4.4, 0), and (6, 6, -0.3), respectively. Solid lines are fitting curves. (**g−i**) are the simulated total, *x*-directional, and *y*-directional projected phonon density of states at different temperature, respectively. Dashed circles indicate the Cu-dominated peak merging with increasing temperature.

Figure 3a-c demonstrates excellent agreement between measured phonon dispersions and our simulation. and Fig. S6a indicates phonon modes around 4 meV is closely related to the vibration of Cu atoms. While TA modes remain well-defined at elevated temperatures, most optical modes (4-6 meV) exhibit substantial broadening. Some of them even become overdamped, as confirmed by the constant-Q scans depicted in Fig. 3d-f. Notably, even TA modes at 300 K show slowly decaying intensity profiles with characteristic tails, often featuring shoulders corresponding to optical modes. These optical peaks broaden significantly and eventually vanish at 600 K (Fig. 3d-f), indicative of overdamping. This overdamping directly correlates with Cu ion dynamics, as evidenced in Fig. 3a-c, g-i and Supplementary Fig. S5. Cs atoms also contribute to the anharmonicity as "rattler" but most of them locate above 4 meV while Cu atoms contribute to the flat optical phonon bands (Supplementary Fig. S6). The density of states (DOS) evolution reveals that discrete peaks between 4-6 meV at 300 K coalesce into a broad feature at 600 K. Projected partial DOS analysis attributes this primarily to Cu atoms: while Cs ions (rattler) exhibit minimal temperature-dependent DOS changes near 4 meV (Supplementary Fig. S5a), and I ions show marginal variations (Supplementary Fig. S5b-c), Cu-derived peaks within 4-6 meV disappear completely with heating, indicating Cu contributes most to the disappearance of total DOS peaks within 4-6 meV in Fig. 3g. As mentioned above, Cu ions hopping is directionally oriented, leading to distinct temperature-dependent DOS evolution along different directions. As evidenced in Fig. 3h-i, DOS of Cu along the *x*-direction exhibits most pronounced DOS changes upon heating, followed by that along the *y*-direction. Along the *z*-direction, peaks broaden without disappearance at higher temperatures. This directional preference aligns with preferential hopping along the *xy*-plane MID pathway, while *z*-direction migration remains infrequent. Crucially, the broadened 4-6 meV peaks at elevated temperatures correspond directly to optical modes exhibiting large phonon linewidths and overdamping (Fig. 3d). We therefore establish that Cu ion hopping constitutes the primary source of enhanced lattice anharmonicity at high temperatures.

**Glass-like thermal conductivity**

Phonon lifetime analysis further corroborates the extreme anharmonicity. As shown in Fig. 4a, b, most phonon lifetimes fall within the wave-like tunneling regime, indicating dominant wave-like phonon behavior. Optical modes near 4 meV reside in the overdamped region, especially the results at 600 K. Considering Cu ions do not hop at 300 K, we extract force constants to describe the phonon properties. The calculation provides clear distribution of phonon lifetimes across the entire phonon energy range. Three-phonon (3ph) scattering confines numerous optical modes to wave-like regions, while higher-energy optical modes exhibit overdamping. Crucially, incorporating four-phonon (4ph) scattering significantly reduces phonon lifetimes, shifting more 4 meV modes into the overdamped regime at 600 K. Notably, the 600 K phonon lifetime serves as an indicative reference for overdamping estimation at elevated temperatures, as Cu ion hopping invalidates the Boltzmann transport equation (BTE) for precise lifetime prediction under these conditions.

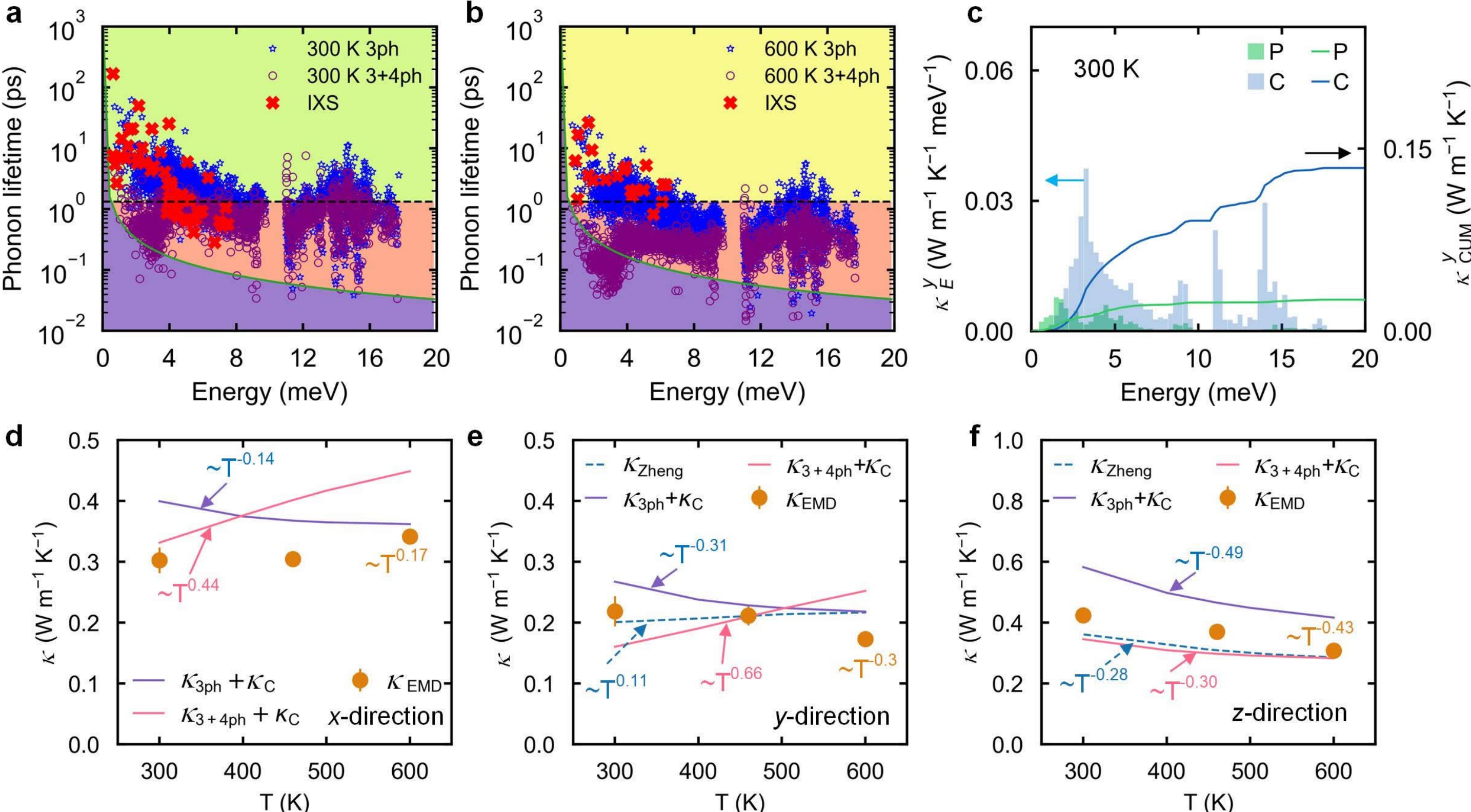

**Fig. 4 | Phonon lifetime and thermal conductivity of $CsCu_2I_3$**. (**a**) and (**b**) are the calculated energy-dependent phonon lifetime at 300 K and 600 K. Blue stars and purple circles are the results of three-phonon (3ph) and that considering the four-phonon (3+4ph), respectively. The red crosses are measured by inelastic X-ray scattering (IXS). The black dashed line represents the Wigner limit $1/\Delta\omega_{ave}$ while the green solid line indicates the Ioffe-Regel limit $1/\omega$ in time, respectively. Phonons exhibit particle-like behavior above the Wigner limit, wave-like behavior between the Wigner limit and the Ioffe-Regel limit, and become overdamped below the Ioffe-Regel limit. **c** Energy-dependent particle-like ($\kappa_P$) and wave-like ($\kappa_C$) thermal conductivities after considering 3+4ph processes at 300 K with their cumulative results along *y*-direction. (**d**-**f**) are the anisotropic total thermal conductivity ($\kappa_P+\kappa_C$), including the results

of $\kappa_{3ph}+\kappa_c$, $\kappa_{3+4ph}+\kappa_c$, and equilibrium molecular dynamics $\kappa_{EMD}$ along *x*-, *y*- and *z*-direction. Dashed lines are results from Ref. 23.

Supplementary Table S5 list the group velocities of $CsCu_2I_3$ from IXS. The value (~2.63 km $s^{-1}$) along $\Gamma$−Y direction is relatively higher compared to systems with ultralow particle-like thermal conductivity ($\kappa_p$), such as ~1.36, ~1.06, and ~1.32 km $s^{-1}$ of $CsPbBr_3$[24], $CsSnI_3$[25], and $Tl_3VSe_4$[26], respectively. Despite this, extreme anharmonicity suppresses $\kappa_p$ to extremely low values: ~0.05, ~0.03, and ~0.13 W $m^{-1}$ $K^{-1}$ along *x*-, *y*-, and *z*-directions, respectively (Fig. 4c and Supplementary Fig. S7), confirming that anharmonicity, not group velocity, controls thermal transport in $CsCu_2I_3$. Consequently, wave-like thermal conductivity ($\kappa_c$) dominates with Cu-related 4 meV modes shown in Fig. S6a contributing ~63% of $\kappa_c$, similar as that along *x*- and *z*-directions.

The hopping motion of Cu ions, particularly at elevated temperatures, contributes significantly to the glass-like temperature dependence of $\kappa$. Initially, we calculated the temperature-dependent thermal conductivity using force constants fixed at 300 K. Under this approximation, $\kappa_{3ph}+\kappa_c$ decreases with increasing temperature along all three directions. Nevertheless, the temperature dependence is notably weaker than the conventional $\kappa \propto T^{-1}$ scaling predicted by dominant 3ph scattering. Specifically, we find $\kappa_x \propto T^{-0.14}$, $\kappa_y \propto T^{-0.31}$, and $\kappa_z \propto T^{-0.49}$ (Fig. 4d-f). Inclusion of 4ph process reveals anomalous, glass-like thermal conductivity behavior ($\kappa$ increasing with T) along the *x*- and *y*-directions, characterized by $\kappa_x \propto T^{0.44}$ and $\kappa_y \propto T^{0.66}$. For comparison, Zheng et al. employed the self-consistent phonon approximation to include temperature effects on force constants, and also reported the glass-like behavior along *y*-direction ($\kappa_y \propto T^{0.13}$), albeit with a weaker relationship than our prediction. In contrast, considering 4ph, our result of $\kappa_z \propto T^{-0.3}$ closely matches Zheng et al.s' of $\kappa_z \propto T^{-0.28}$, suggesting negligible impact of higher-order phonon interactions on $\kappa_c$ along the *z*-direction. However, 4ph significantly reduces the magnitude of $\kappa_p$, causing an apparent decrease of ~33%. At higher temperatures, where Cu ion hopping becomes prominent, phonon lifetime enters the overdamped regime, rendering the standard phonon gas model invalid. Therefore, we employed equilibrium molecular dynamics (EMD) simulations combined with the Green-Kubo method to obtain a more comprehensive description of $\kappa$ (Fig. 4d-f). EMD results show $\kappa_x \propto T^{0.17}$, consistent with the highest Cu ionic diffusivity along *x*-direction. However, $\kappa_y$ and $\kappa_z$ still exhibit decreasing trends ($\kappa_y \propto T^{-0.3}$ and $\kappa_z \propto T^{-0.43}$), differing from the predictions of $\kappa_{3+4ph}$+$\kappa_{C}$.

Analysis of the contributions reveals that vibrational transport ($\kappa_{V}$) dominates the total thermal conductivity (Supplementary Fig. S8). Crucially, the cross-interaction between ionic migration and atomic vibrational ($\kappa_{CROSS}$) along the *x*-direction drives the observed glass-like behavior, while $\kappa_{V}$ itself shows only weak temperature dependence. This indicates that in incipient ionic conductors like this, glass-like $\kappa$ is likely to emerge along the direction of facile ion transport, with $\kappa_{CROSS}$ being the primary contributor to the positive temperature dependence. Collectively, these findings underscore the critical role of Cu ion hopping in inducing glass-like $\kappa$.

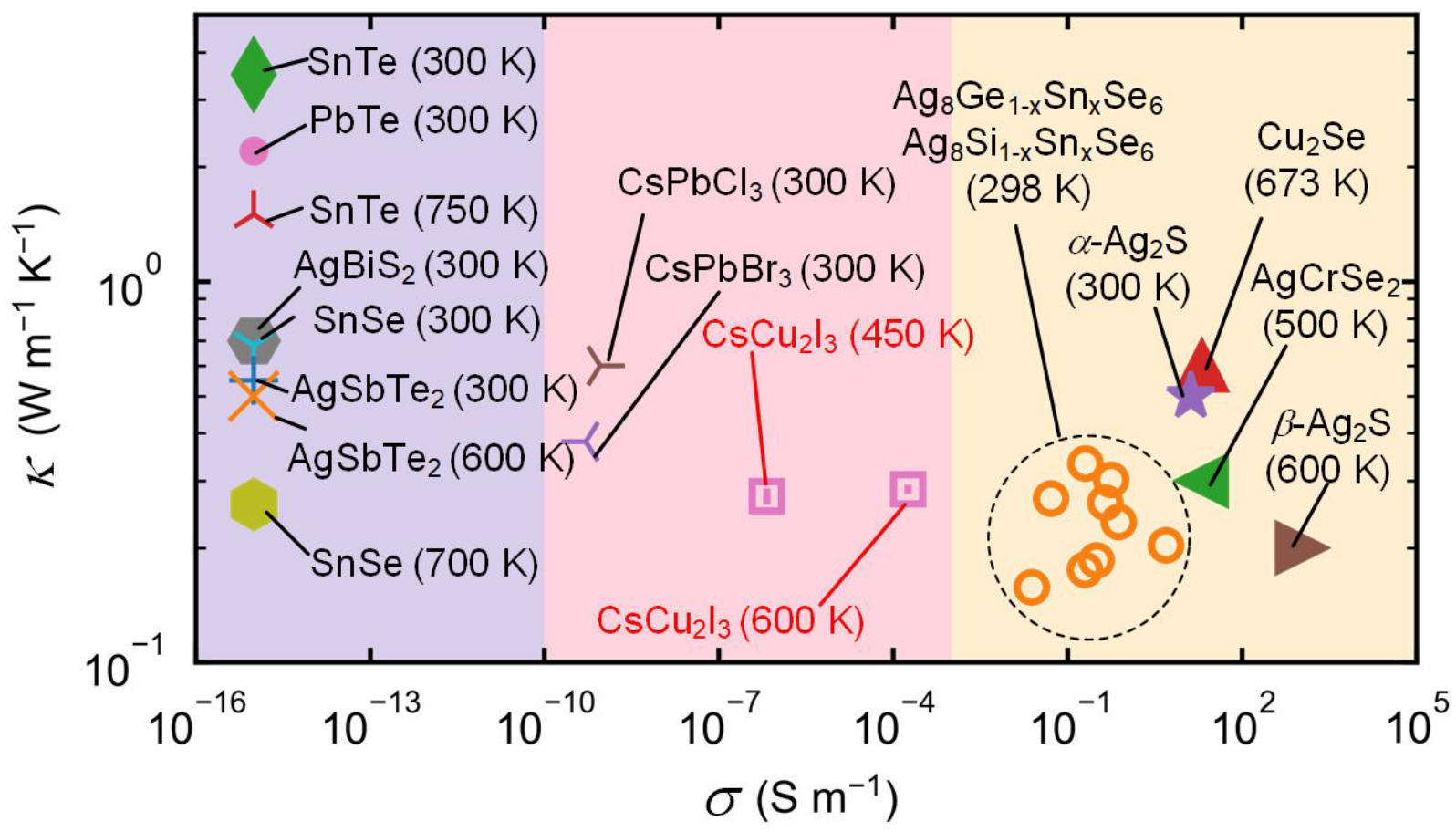

**Fig. 5 | Comparison of ionic transport and thermal conductivity between typical conductors**, showing no direct relationship between ionic and phonon conductivity. Light blue, pink, and green shadows indicate the electronic, incipient, and superionic conductors. Data are from refs. [11,27–39]. Here, an ionic mobility ($\sigma$) of 1×$10^{-16}$ S $m^{-1}$ suggests no ionic conduction.

Ionic transport contributes to the glass-like $\kappa$ of $CsCu_2I_3$, yet $\kappa$ maintains consistent magnitude (~0.30 − ~0.34 W $m^{-1}$ $K^{-1}$) along the *y*-direction (Fig. 5). Despite significantly enhanced ionic conductivity at elevated temperatures, the minimal $\kappa$ variation confirms that this ultra low $\kappa$ stems primarily from extreme anharmonicity induced by ionic vibrations rather than ionic mobility. Figure 5 summaries the ionic transport and thermal conductivity between typical conductors. In electronic conductors with negligible ionic conductivity (e.g., SnTe[31] and PbTe[27] at 300 K), thermal conductivity is generally higher than that of incipient and superionic conductors and superionic conductors. This is primarily due to the limited phonon anharmonicity (including phonon softening and broadening) induced by atomic vibrations. In superionic conductors (e.g., $CuCrSe_2$[5] and $AgCrSe_2$[10]), long-range ionic migration can selectively disrupt transverse phonon modes, leading to material degradation despite glass-like $\kappa$. However, when ions migrate with only short-range displament, such as in $CsCu_2I_3$ and $Cu_{12}Sb_4S_{13}$[8], giant or extreme anharmonicity beyond rattling occurs in systems, yielding glassy-$\kappa$. Notably, high temperature-enhanced anharmonicity can further reduce thermal conductivity of electronic conductors (e.g., SnSe at 750 K[31] in Fig. 5), bringing their $\kappa$ down to levels comparable with incipient and superionic conductors. This demonstrates that high ionic conductivity is not a prerequisite for achieving ultralow $\kappa$. In summary, incipient ionic conductors exhibit anharmonicity levels comparable to superionic conductors while maintaining better stability. Moreover, unlike electronic conductors, they retain ultralow thermal conductivity across a broad moderate temperature range.

## Conclusion

This work identifies $CsCu_2I_3$ as a new incipient ionic conductor. Single crystal synchrotron X-ray diffraction combined with MEM analysis reveals the constrained migration of Cu ions, which preferentially undergo nearest-neighbor exchange between $CuI_4$ tetrahedra along the MID pathway at elevated temperatures. Machine-learning potential-based MD further confirms that this limited migration occurs predominantly along the *x*-direction via the MID pathway. While this Cu ion migration does not disrupt TA modes, it significantly contributes to the observed extreme anharmonicity. Consequently, the motion of Cu ions plays a crucial role in the overdamped optical modes measured by IXS. Phonon modes associated with Cu motion exhibit a distinct propagating character. Furthermore, the mobile Cu ions at high temperatures induce glass-like thermal conductivity by facilitating strong cross-coupling between vibrational (phonon) and ionic-convective heat transport. However, the contribution of mobile Cu ions to the overall change in thermal conductivity is limited. These results propose the Cu/Ag alkali-metal halides as new promissing incipient ionic conductors to simultaneously maintain both glassy $\kappa$ and high stability, reveal the thermal transport mechanism via ion migration constraints, and pave an effective pathway toward ultralow thermal conductivity in ionic conductors.

## Methods

### Sample synthesis

The single crystal $CsCu_2I_3$ was grown by the modified ITC method as shown in our previous work.[40] CsI and CuI in a 1:2 molar ratio were fully dissolved in an OA/DMF (0.8 mL/5 mL) solvent mixture with vigorous stirring for half an hour at room temperature under ambient conditions. In order to remove undesirably undissolved precursors or contaminants, then the resultant greenish solution was filtered into a glass vial using a 0.22 μm poly(tetrafluoroethylene) (PTFE) filter. After that, we quickly sealed the glass vial with cover and placed it in an oil bath. The heating rate was set as 2 °C $h^{-1}$ until the dark brown solution crystallized one or two small $CsCu_2I_3$ seeds in the vials (75 °C). Afterward, the temperature was maintained at 75 °C to promote continuous growth of $CsCu_2I_3$ seeds for 72 h.

**Single-crystal synchrotron X-ray diffraction.**

The high-resolution single-crystal synchrotron X-ray diffraction was conducted at BL02B1 in SPring-8.[41] The photon energy of 50.00 keV with a Pilatus3 X CdTe (P3) detector was utilized, which recently has been regarded as a good tool to achieve extremely high quality for the electron density data.[42] The experiment was conducted at 100, 200, 300, 400 and 500 K with wavelengths of 0.2481 Å. The CrysAlisPro[43] was used to convert the collected frames into Esperanto format and then output the reconstructed files. This reconstruction included the application of an absorption correction using the SCALE3 ABSPACK scaling algorithm. Unwrapped images as shown in Fig. S1 were also based on the CrysAlisPro. The refinement was performed in Olex2[44], where the structures were solved with ShlexT[45]. The detailed structure refinement was conducted by JANA2020[46]. Details of the single-crystal diffraction are shown in Table. S1-S3. The high-quality single crystal synchrotron data provide confidence for the next MEM density analysis.

**MEM density analysis**

We extracted the structure factors from the refined high quality SCSXRD data by Jana2020[46] and then output the file for MEM density analysis with Dysnomia[47]. The Limited-memory BFGS (L-BFGS) algorithm implemented in Dysnomia was chosen for the calculations. The conventional cell of $CsCu_2I_3$ was divided into pixels with the resolution ~0.05 Å. The entropy was defined as $S=-\sum_{k=1}^{N}\rho_k \ln\left(\frac{\rho_k}{\tau_k}\right)$, where $\rho_k$ is the normalized density at position in 3D gridded space and $\tau_k$ is the normalized density derived from prior information. We chose linear combination of generalized constraints with weightings to maximized entropy *S*. The constraints *C* can be described as[47]

$$
\begin{aligned}
C &= \sum_n \lambda_n C_n \\
C_n &= \frac{1}{(N_F + N_G) M_n(\text{ Gauss })}\left[\sum_{j=1}^{N_F} \omega_j \left(\left|\Delta F_j\right|\right)^n + \sum_{j=1}^{G_F} \omega_j \left(\left|\Delta G_j\right|\right)^n\right] - C_{w_n}
\end{aligned}
\qquad (1)
$$

where the $\lambda_n$ is the relative weight of $C_n$. The $\omega_j$ is the weighting factor and $C_{w_n}$ is the criterion for convergence. The $\Delta F_j$ and $\Delta G_j$ are $F$ and $G$ constrain, respectively. $M_n$ is $n$th order central moment of Gaussian distribution. The MEM calculation results are shown in Table. S4.

**Inelastic X-ray scattering**

The BL43LXU beamline at SPring-8 was chosen to conduct inelastic X-ray scattering measurements, where 21.7 keV incident energy at 300 K, 460 K and 600 K. The measurements were performed with a resolution of 1.5 meV and a momentum resolution of 0.05 with units of reciprocal lattice (2π/a). The (010) surface of a high-quality $CsCu_2I_3$ single crystal, approximately 8×1 mm in size as shown in Fig. S10, was oriented perpendicular to the X-ray beam. The (7, 7, 0), (0,8,0), (4, 4, 0) and (6, 6, 0) zones were chosen to measure the phonon information. The dynamical structure factor $S(\mathbf{Q}, E)$ was calculated to guide us design the experiment, which can be described as[48,49]

$$
\begin{aligned}
S(\mathbf{Q},E) \propto \Sigma_s \Sigma_\tau \frac{1}{\omega_{\mathrm{s}}} &\left| \Sigma_d \frac{\overline{b}_d}{\sqrt{M_{\mathrm{d}}}} \exp\left(-W_d\right) \exp\left(\mathrm{i}\mathbf{Q}\cdot\mathbf{r}_d\right)\left(\mathbf{Q}\cdot\mathbf{e}_{ds}\right) \right| \\
&\times \left\langle \mathrm{n}_s + 1 \right\rangle \delta\left(\omega - \omega_{\mathrm{s}}\right) \delta(\mathbf{Q} - \boldsymbol{q} - \tau)
\end{aligned}
\qquad (2)
$$

where $\overline{b}_d$ is neutron scattering length of atom $d$. $\mathbf{Q} = \boldsymbol{k_f} - \boldsymbol{k_i}$ indicates the wave vector transfer, where $\boldsymbol{k_f}$ and $\boldsymbol{k_i}$ are final and incident wave vector of scattered particle. $\boldsymbol{q}$ is phonon wave vector, $\omega_s$ is the eigenvalue of the phonon corresponding to the branch index $s$, $\tau$ is the reciprocal lattice vector, $d$ the atom index in the unit cell, $\mathbf{r}_d$ the atom position, $W_d$ the corresponding Debye-Waller factor, $\mathbf{e}_{ds}$ the phonon eigenvectors. The second-order force constants at 300 K, which are described as in Machine-learning potential based calculation method part, was used to obtain $S(\mathbf{Q}, E)$. In order to acquire the phonon shift and phonon linewidth, we applied the damped harmonic oscillator (DHO) model convoluted with resolution function[50]

$$
I(E) = R * \left( A \frac{4\gamma E}{n(E)\pi} \frac{1}{\left(E - E_C\right)^2 + (2\gamma E)^2} \right) \qquad (3)
$$

where $E_c$, $2\gamma$, $n(E)$ and $A$ are effective frequency of the mode, phonon linewidth (full-width at half-maximum), Bose-

Einstein distribution function at phonon energy transfer $E$ [$n(E) = (e^{E/k_BT} - 1)^{-1}$] and amplitude, respectively.

**First-principle calculation**

We conducted first-principles calculation with Vienna Ab initio Simulation Package (VASP)[51]. The exchange-correlation function PBEsol was chosen to relax the lattice, with kinetic energy cutoff 700 eV and electronic energy tolerance $10^{-6}$ eV. The K-points were chosen as 6 × 6 × 8. The acquired lattice constants $a$ = 10.04 Å, $b$ = 12.94 Å and $c$ = 6.05 Å are close to the experimental $a$ = 10.48 Å, $b$ = 13.04 Å and $c$ = 6.06 Å at 100 K.

**Machine-learning potential based calculation**

The machine learning interaction potential moment tensor potential (MTP) developed by Shapeev et al.[52] was chosen to conduct molecular dynamics (MD) simulation. Ab initio molecular dynamics calculations were performed with 2 × 2 × 2 supercell of primitive cell at 300, 450, and 600 K to provide the training database. In this process, we also scaled the volume range from 0.95 to 1.05 at 300 , 450, and 600 K to provide more configurations for MTP training. The energy cutoff and energy tolerance were set as 600 eV and $10^{-6}$ eV, respectively. The force tolerance was set as $10^{-3}$ eV Å$^{-1}$. The K-point was set as $\Gamma$-only. We chose exchange-correlation function PBEsol to run each simulation for 1200 steps with a timestep of 1 *fs*/step, with a Nosé-Hoover thermostat controlling temperature. The 3486 configurations were chosen as training data to train MTP with a maximum level of 22. The minimum and maximum cutoff radii for the MTP were chosen to be 2.13 Å and 7 Å. The fitting weights of energies, forces, and stresses were set as 1, 0.1, and 0.01 in the training processes, respectively. After that, we conducted the active learning strategy based on the D-optimality[53], as included in MLIP package, to improve the performance of MTP potential. The break threshold ($\gamma_{break}$) and selection threshold ($\gamma_{select}$) were set as 10 and 3. New configurations were selected after independent MD simulation, and then we performed first-principles calculations for these new structures. These configurations were then added to retrain the MTP potential. We performed 13 iterations and a total of 86 configurations were actively selected. For active learning, the MD simulation was run for 1 *ns*. The validation of energy and force is shown in Fig. S11.

Then we conducted MD by LAMMPS[54] at 300, 450, and 600 K with the 3 × 3 × 5 supercell of conventional cell. All systems were equilibrated for 2 ns with *NVT* ensemble and then were switched to *NVE* ensemble to collect data for 2 ns. Each step was set as 2 fs in all simulations. The 200 configurations at 300 K were selected to extract the force constants by HIPHIVE[55]. The second, third and fourth order cutoff distance were set as 8, 6, and 5.2 Å respectively, which converged as shown in Fig. S12a-c. These force constants were used to calculate the thermal conductivity by FOURPHONON[56,57] with a converged q-mesh 13 × 13 × 13 as shown in Fig. S12d. Considering the migration of cooper ions at high temperature, the velocity of Cs, Cu and I atoms was extracted from the trajectories to obtain the phonon density states by velocity autocorrelation function, which is described as $\mathrm{PDOS}(\omega)=\int\left\langle\sum_i v_i(t_0)\cdot v_i(t_0+t)\right\rangle/\left\langle\sum_i v_i(t_0)\cdot v_i(t_0)\right\rangle\exp(-2\pi i\omega t)dt$. The $t$ is correlation time and the <·> indicates the average over all atoms or same elements[58]. The MSD was extracted by MDanalysis[59]. The diffusivity was calculated with $\mathrm{MSD}_{Cu}(t) = 6D_{Cu}t$.[20,60] The dynamical structure factor was calculated from trajectories by

$$\begin{aligned} S(\mathbf{Q},\omega) &= \int\langle\hat{\rho}(\mathbf{r},t)\hat{\rho}(0,0)\rangle\exp(-i(\mathbf{Q}\cdot\mathbf{r}-\omega t))dtd\mathbf{r} \\ &\equiv \int G(\mathbf{r},t)\exp(-i(\mathbf{Q}\cdot\mathbf{r}-\omega t))dtd\mathbf{r} \end{aligned} \tag{4}$$

where $G(\mathbf{r},t)$ is regarded as Von-Hove Function and the dynamical structure factor is its time- and space- Fourier transformation.[49,61] The $G(\mathbf{r},t)$ can be described as $G(\mathbf{r},t)=\Sigma_i^N\Sigma_j^N b_i b_j\int\delta\left(\mathbf{r}-\left(\mathbf{r}_\mathbf{i}\left(t+t'\right)-\mathbf{r}_\mathbf{j}\left(t'\right)\right)\right)dt'$, where $\mathbf{r}_i(t)$ is the $i$th atom position at time $t$. The $b_i$ is the scattering length of $i$th atom. In these simulations, we used 10 × 10 × 10 supercell of conventional cell, totally 24,000 atoms. The data collection was conducted with *NVE* ensemble for 50 *ps* after equilibration with *NVT* ensemble for 50 *ps* with a timestep 1 fs. We calculated the $S(\mathbf{Q}, E)$ at 300 K and 600 K along (7, 7-K, 0), (H, 4, 0) and (0, 8, -2+L). The momentum and energy resolutions based on such simulation are ~0.15 Å and ~0.41 meV, respectively.

The thermal conductivity is calculated by MTP-based EMD, which relies on the Green-Kubo method. The heat flux $\mathbf{J}(t)$ is described as[6]

$$\mathbf{J}(t)=\sum_i \mathbf{v}_i\varepsilon_i+\frac{1}{2}\sum_{ij,i\neq j}\mathbf{r}_{ij}\left(\mathbf{F}_{ij}\cdot\mathbf{v}_{ij}\right) \tag{5}$$

where $\mathbf{v}_i$ and $\varepsilon_i$ are the velocity and energy of the $i$th atom, respectively. $\mathbf{F}_{ij}$ and $\mathbf{r}_{ij}$ are the force and distance between $i$th and $j$th atoms, respectively. The first term is convection heat flux $\mathbf{J}_{\mathrm{conv}}(t)$, which is related to the ionic transport. The second term is conduction heat flux $\mathbf{J}_{\mathrm{virial}}(t)$, which is from the lattice vibrations in solids. The total lattice thermal conductivity $\kappa_{\mathrm{latt}}$ is expressed as below[62]:

$$\kappa_{\text{latt}} = \frac{1}{3k_B T^2 V}\left[\int_0^t \langle J_{virial}(0)\cdot J_{virial}(t)\rangle dt + \int_0^t \langle J_{conv}(0)\cdot J_{conv}(t)\rangle dt + 2\int_0^t \langle J_{virial}(0)\cdot J_{conv}(t)\rangle dt\right] \quad (6)$$

The first term in the square bracket is contribution from atomic vibration $\kappa_{\text{V}}$. The second term is the contribution from the non-phononic convection of migration atoms $\kappa_{\text{CONV}}$ and the third term is the contributions from the cross interactions between ionic migration and atomic vibrational $\kappa_{\text{CROSS}}$. The 4 × 3 × 5 supercell of conventional cells were used in all calculation. All systems were firstly equilibrated with *NVT* ensemble and then we collected data with *NVE* ensemble. The data collection for the heat flux calculation is 25 ns with a time step 1 fs, which were separated to 25 iterations.

## Acknowledgments

We acknowledge Yuiga Nakamura at BL02XU@SPring-8 and Jiesheng Hu at School of Chemistry and Chemical Engineering in Beijing Institute of Technology for discussion. The work at Beijing Institute of Technology is supported by the National Key R&D Program of China (2021YFA1400300), the Open Fund of the China Spallation Neutron Source Songshan Lake Science City (KFKT2023A07), Beijing National Laboratory for Condensed Matter Physics (2023BNLCMPKF003). The work at Beihang University thank the support of National Natural Science Foundation of China (No. 52125205, 52250398, 52192614 and 52203307), Natural Science Foundation of Beijing Municipality (2222088 and L223006), Shenzhen Science and Technology Program (Grant No. KQTD20170810105439418) and the Fundamental Research Funds for the Central Universities. The authors also thank Xuzhou B&C Chemical Co. Ltd for providing the photoresist (HTA116, HTA112, B&C Chemicals) used in our work. Part of this research was performed at BL43LXU (proposal No. 2022B1379) and BL02XU (proposal No. 2023B2064) at the SPring-8, Japan, under a user program.

## Author contributions

Y.L., B.W., and J.H. designed the project. Y.L., C.L., D.I., T.M. and A.B. carried out the IXS experiment analyzed the data. Y.L. and J.H. performed the simulation. C.L. and C.P. prepared the single crystal. Y.L., B.W, X.J., and J.H. discussed this work. Y.L. and B.W. wrote the original manuscript. C.P. and J.H. supervised the project. All the authors reviewed and edited the manuscript.

## Data and materials availability

All data related to the conclusion are present in the paper and the Supplementary Materials. The data related to this paper can be requested from the authors.

## Competing interests

The authors declare that they have no known competing financial interests or personal relationships that could have appeared to influence the work reported in this paper.